\documentclass[preprint2]{aastex}

\def\eg{{e.g.~}}
\def\vc{{Virgo cluster}}
\def\VC{{Virgo Cluster}}
\def\mc{{WHL~J085910.0+294957}}
\def\ccom{{$(g-r)_{com}$}}
\def\cbri{{$(g-r)_{bri}$}}
\def\gpi{{\mbox{\boldmath${P}$}}}

\shorttitle{Small-scale Conformity in the \vc}

\shortauthors{Lee et al.}

\begin{document}

\title{SMALL-SCALE CONFORMITY OF THE VIRGO CLUSTER GALAXIES}

\author{Hye-Ran Lee$^{1,2}$, Joon Hyeop Lee$^{1,2}$, Hyunjin Jeong$^{1,2}$, Byeong-Gon Park$^{1,2}$}
\affil{$^1$ University of Science and Technology, Daejeon 34113, Republic of Korea\\
$^2$ Korea Astronomy and Space Science Institute, Daejeon 34055, Republic of Korea}

\email{hrlee@kasi.re.kr}

\begin{abstract}
We investigate the small-scale conformity in color between bright galaxies and their faint companions in the {\vc}. Cluster member galaxies are spectroscopically determined using the Extended {\VC} Catalog (EVCC) and the Sloan Digital Sky Survey Data Release 12 (SDSS DR12). We find that the luminosity-weighted mean color of faint galaxies depends on the color of adjacent bright galaxy as well as on the cluster-scale environment (gravitational potential index). From this result for the entire area of the {\vc}, it is not distinguishable whether the small-scale conformity is genuine or is artificially produced due to cluster-scale variation of galaxy color. To disentangle this degeneracy, we divide the {\vc} area into three sub-areas so that the cluster-scale environmental dependence is minimized: A1 (central), A2 (intermediate) and A3 (outermost). We find conformity in color between bright galaxies and their faint companions (color-color slope significance $\emph{S}\sim2.73\,\sigma$ and correlation coefficient $\emph{cc}\sim0.50$) in A2, where the cluster-scale environmental dependence is almost negligible. On the other hand, the conformity is not significant or very marginal ($\emph{S}\sim1.75\,\sigma$ and $\emph{cc}\sim0.27$) in A1. The conformity is not significant either in A3 ($\emph{S} \sim1.59 \,\sigma$ and $\emph{cc} \sim0.44$), but the sample size is too small in this area. These results are consistent with a scenario in which the small-scale conformity in a cluster is a vestige of infallen groups and these groups lose conformity as they come closer to the cluster center.       

\end{abstract}


\section{INTRODUCTION}

The evolution of galaxies is known to be significantly affected by their environment. For example, the fraction of early-type galaxies increases along local number density \citep[\eg][]{dre80, bal06, par07, lee10}, while the star formation rate (SFR) and gas content of a galaxy decrease \citep[\eg][]{kau04,pog08}. As a result, the galaxies in dense environments tend to have early-type morphologies, red colors, and low SFRs. Such trends are generally thought to be related to the acceleration of galaxy evolution by certain mechanisms, such as galaxy interactions in high-density environments.

Galaxy evolution by direct tidal interactions between galaxies is known to be most active in a galaxy group \citep{per09, alo12, pau14}. Even if galaxies in a group do not interact with one another directly, they evolve within a single halo and thus share their small-scale environment. As a result, the properties of central bright galaxies tend to be closely related to those of their satellites, in their morphologies, colors, and SFRs. This relationship is called ``galactic conformity'', which was first introduced by \citet{wei06}. After the first discovery, many subsequent studies found additional evidence supporting this relationship. \citet{ann08}, for example, reported that the early-type host galaxies tend to have early-type satellites and \citet{kau10} showed that the colors of satellites are related to the color of their central galaxy. It has also been reported in several studies that satellite quenching efficiency is connected to the SFR of a central galaxy \citep{kau13, phi14, kaw15, kno15, phi15}. According to \citet{wan15}, low-mass galaxies around HI-rich central galaxies tend to have extended gas reservoirs.  

In case of a galaxy cluster, various mechanisms are known to considerably influence galaxy evolution. For example, cluster galaxies are affected by interaction with cluster potential \citep{mer84, gne03}, harassment \citep{moo96, moo99}, galaxy-galaxy hydrodynamic interaction \citep{ph09}, strangulation \citep{lar80, bek02} and ram-pressure stripping \citep{gun72, qui00}. However, direct interaction between galaxies is considered to be hardly possible because galaxies have very high relative velocities in a massive cluster. For this reason, small-scale conformity was less expected to be found in cluster environments, and it has been rarely studied whether such small-scale conformity also exists in galaxy clusters.

However, although galactic conformity may be hardly born in cluster environments, galaxy clusters are known to grow by merging many galaxy groups. Thus, if small-scale conformity between bright galaxies and their faint companions remains for some time after a galaxy group falls into a cluster, the conformity may be detected even in cluster environments. \citet{lee14} addressed this issue for {\mc}, a cluster at redshift 0.3, based on their observation data using the 2.1 m Otto Struve Telescope and the Camera for Quasars in EArly uNiverse (CQUEAN) at McDonald Observatory. They found that the weighted mean color of companions tends to be marginally redder as their bright host galaxy is redder, whereas it depends neither on the luminosity of the host galaxy nor on the cluster-centric distance. To explain the result, they suggested three scenarios regarding the origin of the small-scale conformity in a cluster: tidal dwarfs, dwarf capture, and vestige of infallen groups. However, the pilot study of \citet{lee14} has two major limitations. First, their member selection mostly relied on photometric methods, the reliability of which is not so high. Since the signal of small-scale conformity in \citet{lee14} is not sufficiently significant ($\sim2.2\,\sigma$ significance), the cluster member selection needs to be improved to get results with higher reliability. Second, their observation covered only the central area of the target cluster ($< 1$ Mpc in diameter). To determine the most reasonable scenario about the origin of the small-scale conformity in a cluster, studies covering more extended areas of clusters are necessary.  

The goals of this work are to find the small-scale conformity in another cluster with better member selection and more extended spatial coverage, and to examine the three scenarios. To achieve these goals, we investigate the Virgo cluster using the Extended Virgo Cluster Catalog (EVCC; \citealt{kim14}) and the Sloan Digital Sky Survey Data Release 12 (SDSS DR12; \citealt{ala15}). The structure of this paper is as follows. Section~\ref{data} gives information regarding the data set used in this study. Section~\ref{anal} describes the analysis method and defines additional parameters. Section~\ref{result} presents the results, and their implications are discussed in Section~\ref{dc}. The paper is concluded in Section~\ref{cc}.   

\begin{figure*}[t]
\centering
\includegraphics[angle=0, width=0.24\textwidth]{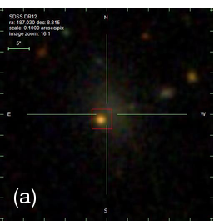}
\includegraphics[angle=0, width=0.24\textwidth]{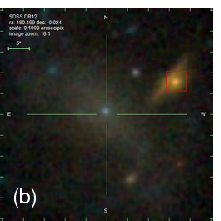}
\includegraphics[angle=0, width=0.24\textwidth]{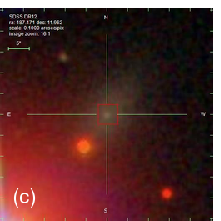}
\includegraphics[angle=0, width=0.24\textwidth]{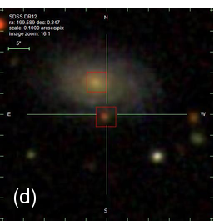}
\caption{Some examples of problematic cases in the radial velocity matching. The center of the green lines indicates the photometric center of each galaxy and the red square shows the target position of the SDSS DR12 spectroscopy. (a) The radial velocity of the galaxy is obtained from a foreground star. (b) The radial velocity of the galaxy is obtained from a nearby galaxy. (c) The spectrum of the target galaxy is probably contaminated by a bright and close star. (d) The target is mismatched due to bad detection.}   
\label{image}
\end{figure*}

\section{DATA}\label{data}

This paper is based on the EVCC (\citealt{kim14}), which is an extended version of the {\VC} Catalog (VCC; \citealt{bin85}), and was complemented using the SDSS DR7 \citep{aba09}. The EVCC galaxies were selected from a wide region that is 5.2 times larger than the field of the VCC. \citet{kim14} improved the morphological classification of the EVCC galaxies by visually checking the multi-band images from the SDSS. However, VCC galaxies fainter than r = 17.7 are not included in the EVCC due to the observational limit of the SDSS spectroscopy. A total of 1,589 galaxies with radial velocities less than 3000 km s$^{-1}$ are selected as Virgo cluster members in the EVCC. Among them, only 913 EVCC galaxies are VCC objects, and the others are newly added by \citet{kim14}. 

The EVCC provides much useful information, such as heliocentric radial velocity and cluster membership, as well as various photometric and structural parameters measured using the \emph{Source Extractor} \citep{ber96}, and manually classified morphology. In this paper, coordinates, magnitudes, membership and morphological information of the Virgo galaxies are retrieved from the EVCC. The spectroscopically determined cluster membership information in the EVCC is very helpful to enhance the reliability of our study. The magnitudes are corrected using the foreground galactic extinction map from \citet{sch98} and the reddening law of \citet{car89} by \citet{kim14}. To calculate absolute magnitude, we adopt the distance modulus $m - M = 31.1$ (\citealt{jer04}; \citealt{mei07}).

\begin{figure}[!t]
\includegraphics[angle=0, width=0.5\textwidth]{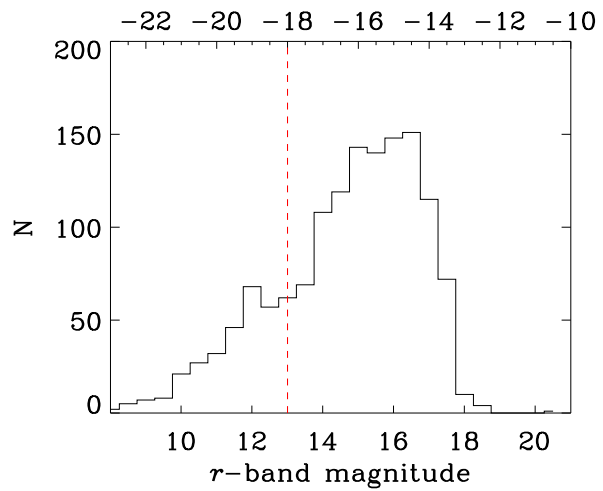}
\caption{Galaxy luminosity function in the $r$-band for the Virgo cluster. The cluster members are divided into bright galaxies and faint ones by a simple cut at -18 mag in the $r$-band absolute magnitude (red dashed line). 199 members are selected as bright galaxies.}
\label{lf}
\end{figure}

\citet{kim14} used the SDSS DR7 for the EVCC, but now the SDSS DR12 is available, which provides more reliable measurements for various parameters by using improved pipelines. Thus, in this paper, we complement the radial velocities of the galaxies using the velocity information from the SDSS DR12 to achieve more convincing membership determination of cluster galaxies. First, we matched the EVCC galaxies with the SDSS DR12 galaxies based on their coordinates. For galaxies without SDSS radial velocities, velocity information from the NASA/IPAC Extragalactic Database\footnote{http://ned.ipac.caltech.edu/} was used. After that, the galaxies whose radial velocities seem to be problematic were excluded by visually inspecting the SDSS DR12 images, some examples of which are shown in Figure~\ref{image}. Through this visual check, 72 problematic objects were removed. Among 1,517 galaxies with velocity information, 102 galaxies with radial velocities larger than 3000 km s$^{-1}$ were additionally excluded, because they can hardly be member galaxies of the {\vc}. As a result, among the EVCC galaxies, 1,262 galaxies with SDSS radial velocities and 153 galaxies with NED radial velocities are finally used in this paper.  

\begin{figure*}[!t]
\centering
\includegraphics[angle=0, width=0.9\textwidth]{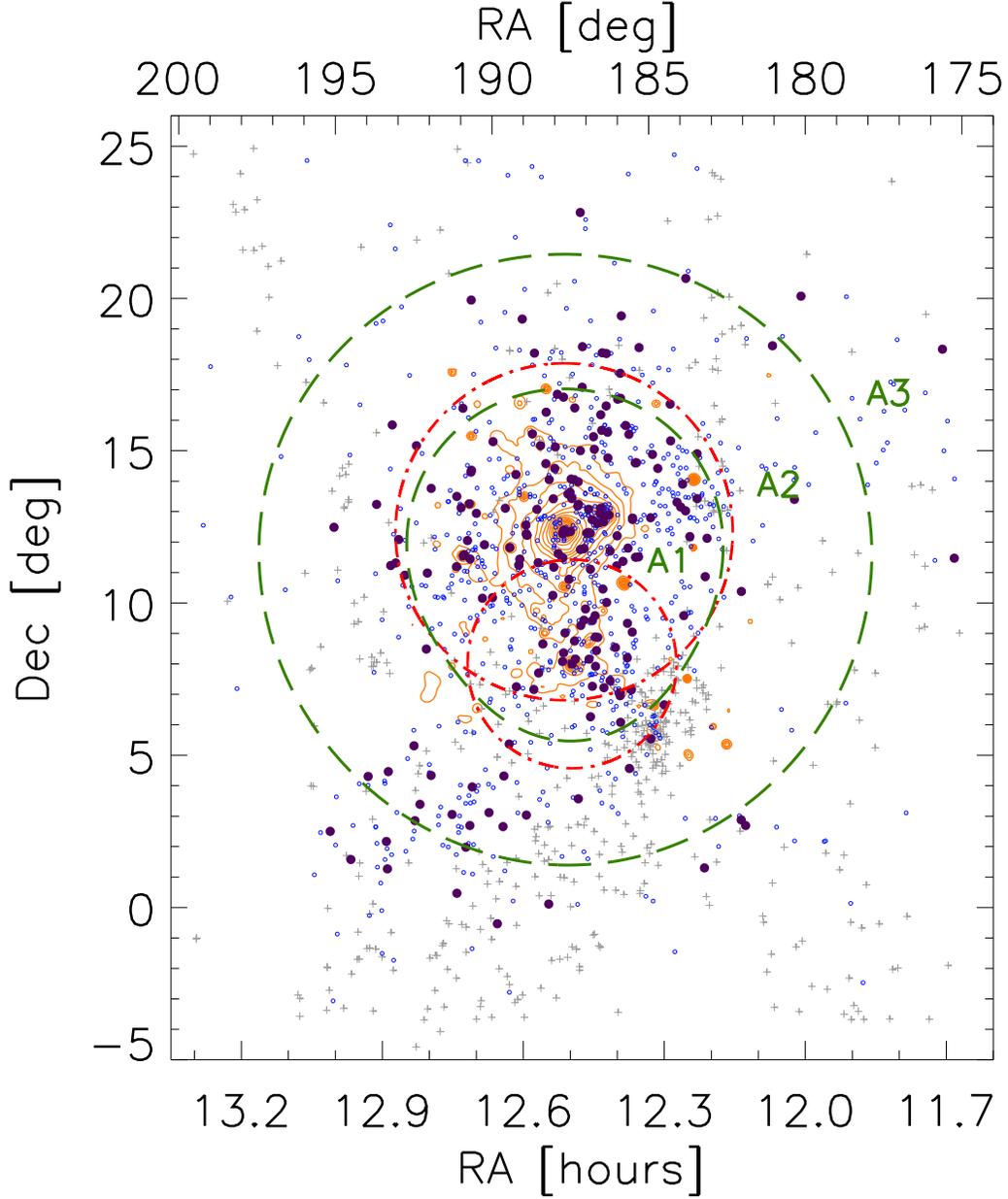}
\caption{Spatial distribution of the finally-selected member galaxies in the {\vc}. Different symbols indicate galaxies with different magnitude ranges and membership classes: genuine bright member (large violet filled circle), genuine faint member (small blue open circle), and possible member (gray cross). The orange solid contours show the X-ray map \citep{boh94}. There are several strong X-ray regions without cluster member counterparts, which are emitted by foreground or background sources. The two red dot-dashed circles are centered on M87 and M49 with their virial radii, respectively. The equi-potential levels are denoted as green long-dashed contours with ${\gpi} = 0.47$ (inner contour) and ${\gpi} = 0.25$ (outer contour), which divide the area into three sub-areas, A1, A2 and A3.}  
\label{gp}
\end{figure*}

\section{ANALYSIS}\label{anal}  

\subsection{Definition of Bright Galaxies and Their Faint Companions}\label{defgal}
The EVCC provides membership information determined using the {\vc} infall model from \citet{pra94}. The provided membership is classified into genuine and possible members based on the distribution over the phase-space diagram in Figure 4 of \citet{kim14}: genuine members within the caustic curve derived from the infall model and possible members out of the caustic curve. Here, we use only 918 genuine members to reduce uncertainty in our analysis.

To investigate small-scale conformity in the {\vc}, we define possible groups within the cluster. We first divide the genuine members into bright galaxies and faint ones with a simple cut at -18 mag in the $r$-band absolute magnitude. The criterion was determined from the bright-end bump in the $r$-band luminosity function as shown in Figure~\ref{lf}. The shape of the luminosity function in the {\vc} is similar to that in {\mc} \citep{lee14}, including the existence of the bump. 

The bright galaxies are defined as the genuine members with $M_r \le -18$. The faint companions of each bright galaxy are defined as the genuine members with $M_r > -18$, which satisfy two additional conditions: within the virial radius of the brht galaxy and velocity dispersion of the expected group. First, the faint galaxies only within the virial radius of a bright galaxy are regarded as its companions in this paper, because the virial radius reflects the boundary to which a galaxy gravitationally influences its companions. \citet{ph09} presented empirical relations between the absolute magnitudes and virial radii of galaxies for early- and late-type galaxies, respectively. Here, the virial radii of bright galaxies were interpolated from those empirical relations. Second, the faint galaxies only within the expected velocity dispersion range of a possible group are regarded as companions of its central bright galaxy. The average velocity dispersions of galaxy groups along the luminosity of group-central galaxies are given by \citet{li12}. By assuming that selected bright galaxies are the central galaxies of possible groups, the expected velocity dispersions of the possible groups are interpolated from Table 1 of \citet{li12}. Through this procedure, a total of 199 possible groups are selected. 

We define the luminosity-weighted mean color of faint companions {\ccom} to be compared with the color of their adjacent bright galaxy {\cbri}. Our definition of {\ccom} is: 

\begin{equation}
(g-r)_{com} = \frac{\sum_{k=1}^{N} (g-r)_{k}W_{L_{k}}}{\sum_{k=1}^{N} W_{L_{k}}}, 
\end{equation}
where $(g-r)_{k}$ is the extinction-corrected color of the $k$-th companion, $N$ is the number of companions, and $W_{L_{k}}$ is the luminosity weight of each faint companion, defined as:

\begin{equation}
W_{L_{k}} = 10^{-0.4({M_r}_{k} + C)},
\end{equation}
where ${M_r}_{k}$ is the $r$-band absolute magnitude of the $k$-th companion and $C$ is an arbitrary constant.

Here, we restrict the sample to the bright galaxies with at least two faint companions for higher statistical reliability. As a result, a total of 165 possible groups are finally selected. The average number of companions in each possible group is 13 and the standard deviation of color indices in a possible group is 0.116 on average.

\begin{figure*}[!t]
\includegraphics[angle=0, width=0.5\textwidth]{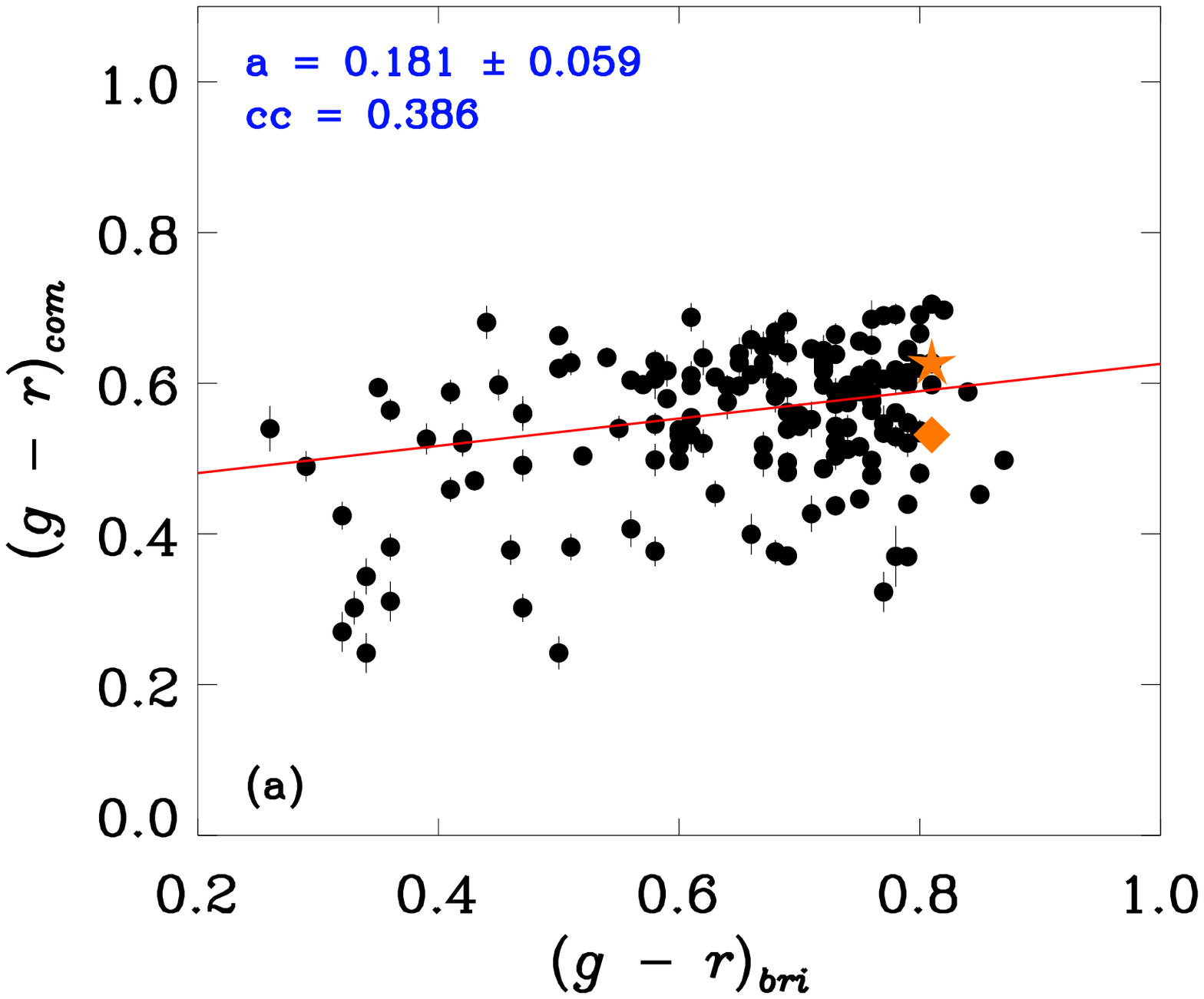}
\includegraphics[angle=0, width=0.5\textwidth]{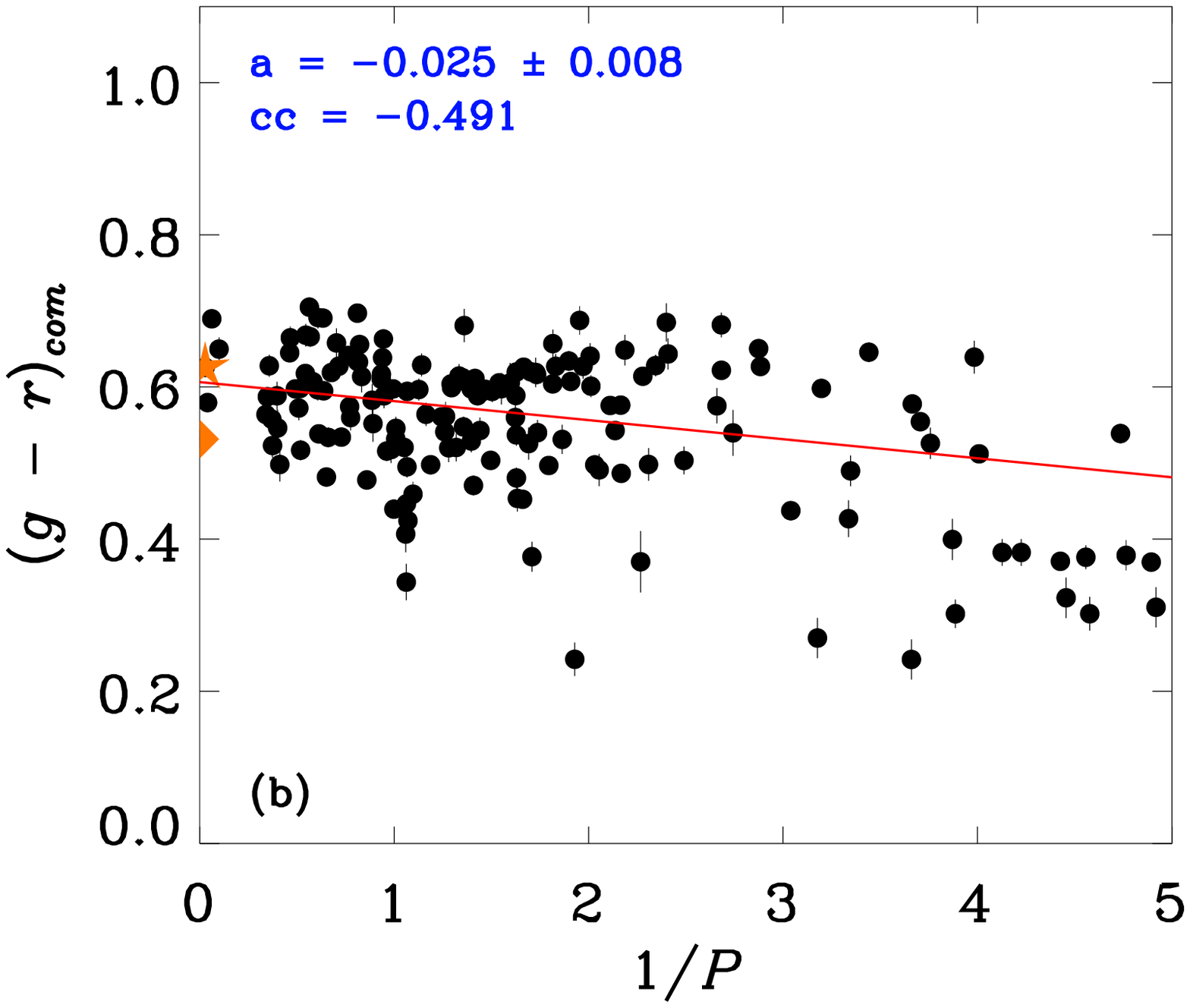}
\caption{The luminosity-weighted mean color of faint companions {\ccom} (a) versus the color of their adjacent bright galaxy {\cbri} and (b) versus the inverse {\gpi}, over the entire region of the Virgo cluster. The star- and diamond-shaped symbols denote M87 and M49, respectively. In each panel, the solid line is the linear least squares fit. The slope of the fit and its bootstrap uncertainty (\emph{a} value), as well as the correlation coefficient (\emph{cc} value), are presented in each panel.}
\label{colcol_full}
\end{figure*}

\subsection{Gravitational Potential Index}\label{pt}
One of the most frequently used parameters to determine how certain quantities depend on a cluster environment is the distance from the cluster center. However, here we define a gravitational potential index ({\gpi}) to quantify the cluster environment instead of the simple cluster-centric distance, because there are two massive galaxies M87 and M49 in the {\vc} central area. Both galaxies are quite dominant in the {\vc} and thought to significantly affect the cluster members. It is the normalized sum of the gravitational potentials from the two galaxies, as follows:  

\begin{equation}
{\gpi}=<\frac{m_{87}}{r_{87}} + \frac{m_{49}}{r_{49}}>,
\end{equation} 
where $m_{87}$ and $m_{49}$ are the masses of M87 and M49 \footnote{The mass of M87 is 4.2 times larger than that of M49 \citep{fer12}.}, respectively, and $r_{87}$ and $r_{49}$ are projected distances from M87 and M49, respectively. For convenience, normalization is carried out so that 1 is the ${\gpi}$ value at the point on the line connecting M87 with M49, where the summed gravitational potential is minimum between the two galaxies.  

In Figure~\ref{gp}, the equi-potential contours are shown: the inner contour at ${\gpi} = 0.47$ and the outer one at ${\gpi} = 0.25$. A smaller {\gpi} value approximately corresponds to a larger cluster-centric distance. In the following plots, we use the inverse {\gpi} parameter (1/{\gpi}), which is a proxy for the cluster-centric distance. A large 1/{\gpi} value approximately indicates a large distance from the cluster center.

\section{RESULTS}\label{result} 

\begin{figure*}[!t]
\centering
\includegraphics[angle=0, width=0.9\textwidth]{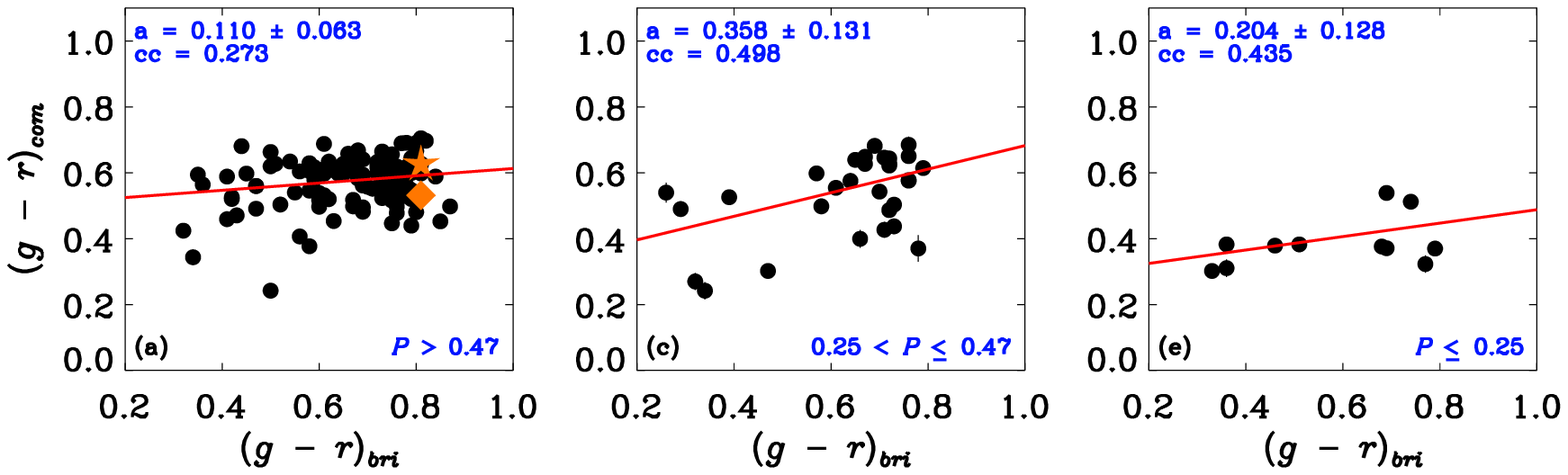}
\includegraphics[angle=0, width=0.9\textwidth]{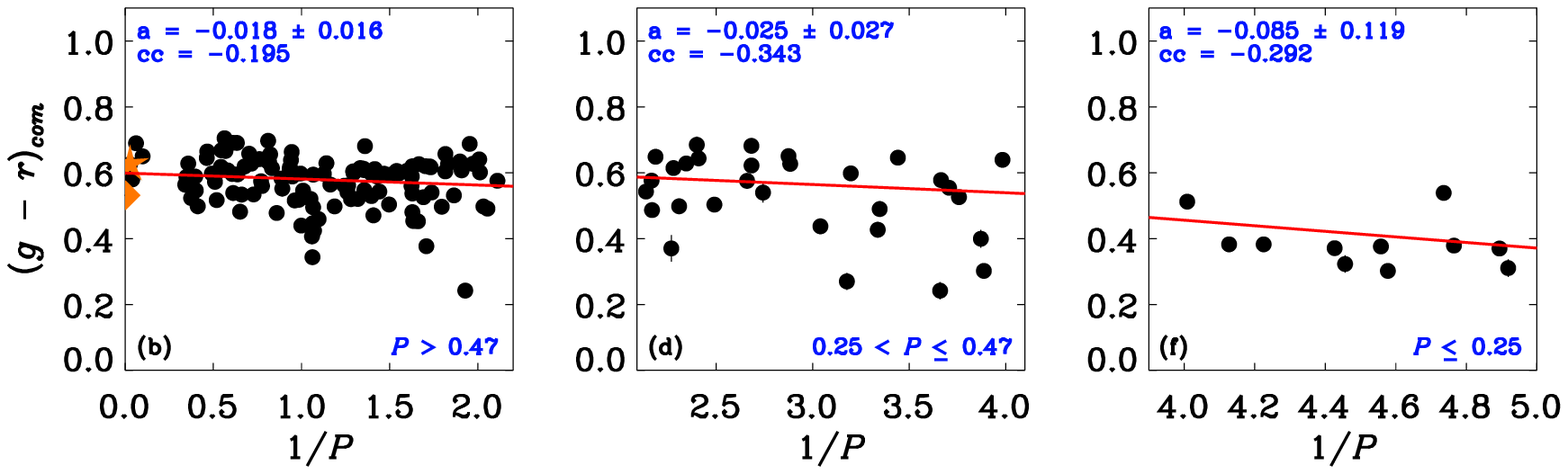}
\caption{The same as Figure~\ref{colcol_full}, but for three sub-areas A1 (${\gpi} > 0.47$: panels a and b), A2 ($0.25 < {\gpi} \le 0.47$: panels c and d), and A3 (${\gpi} \le 0.25$: panels e and f).}
\label{colcol_div}
\end{figure*}

\subsection{Entire area}\label{ea}
Figure~\ref{colcol_full} presents the dependence of the luminosity-weighted mean color of faint companions on the color of their adjacent bright galaxy and on the gravitational potential index. In Figure~\ref{colcol_full}(a), a trend is found that the redder bright galaxies tend to have redder faint companions. For example, {\ccom} around bright galaxies with $(g-r)_{bri} < 0.6$ ranges from 0.2 to 0.7, while that around bright galaxies with $(g-r)_{bri} > 0.6$ ranges from 0.4 to 0.75. In Figure~\ref{colcol_full}(b), however, {\ccom} also depends on {\gpi}, in the sense that faint companions with smaller 1/{\gpi} values (closer to the cluster center) tend to be redder. We estimated a linear least squares fit for each panel, calculating the relation slope (\emph{a}) and its bootstrap uncertainty, as well as the correlation coefficient (\emph{cc}). The two massive galaxies M87 and M49 denoted on the plots are slightly displaced from the linear fits, to the upper and lower directions, respectively, but neither of their displacements is so significant. 

The fitting results are listed in Table~\ref{tres}. Here, it is shown that the slope significance (\emph{S}) is as large as $3.07\,\sigma$ and \emph{cc} is approximately $0.39$ for the {\ccom} versus {\cbri} relation. Meanwhile, {\ccom} also significantly depends on {\gpi}: \emph{S} $\sim 3.13 \,\sigma$ and $\emph{cc} \sim -0.49$. That is, since the {\ccom} is related to both parameters, it is not clear whether the correlation in color between bright galaxies and their faint companions originates from genuine small-scale conformity of possible groups or is artificially produced due to the common environmental dependence of the colors of bright and faint galaxies.

\begin{deluxetable}{rlcr @{$\pm$} lr @{.} lr @{.} l}
\tablenum{1} \tablecolumns{9} \tablecaption{Results of Statistical Tests for Figures~\ref{colcol_full} and ~\ref{colcol_div} \label{tres}} \tablewidth{0pt}
\tablehead{\multicolumn{2}{c}{Figure} & Sample & \multicolumn{2}{c}{Slope $^{(i)}$} & \multicolumn{2}{c}{Significance} & \multicolumn{2}{c}{Correlation} \\
& & Size & \multicolumn{2}{c}{} & \multicolumn{2}{c}{(\emph{S})} & \multicolumn{2}{c}{Coefficient (cc)} }
\startdata
\multicolumn{9}{c}{(Entire Area)} \\
\ref{colcol_full}
& (a) & 165 & $0.181$ & 0.059 & $3$ & 068 & $0$ & 386 \\
& (b) & 165 & $-0.025$ & 0.008 & $-3$ & 125 & $-0$ & 491 \\
\hline
\multicolumn{9}{c}{(Divided Areas)} \\
\ref{colcol_div}
& (a) A1 & 124 & $0.110$ & 0.063 & $1$ & 746 & $0$ & 273 \\
& (b) A1 & 124 & $-0.018$ & 0.016 & $-1$ & 125 & $-0$ & 195 \\
& (c) A2 & 30 & $0.358$ & 0.131 & $2$ & 733 & $0$ & 498 \\
& (d) A2 & 30 & $-0.025$ & 0.027 & $-0$ & 926 & $-0$ & 343 \\
& (e) A3 & 11 & $0.204$ & 0.128 & $1$ & 594 & $0$ & 435 \\
& (f) A3 & 11 & $-0.085$ & 0.119 & $-0$ & 714 & $-0$ & 292 \\
\enddata
\tablecomments{$(i)$ The slope from linear squares fits and their bootstrap uncertainties.}
\end{deluxetable}

\subsection{Divided areas}\label{da}
In the analysis for the entire sample, degeneracy is found between the dependence on {\cbri} and the dependence on the cluster environment. Thus, it is necessary to control the environmental dependence for the investigation of genuine small-scale conformity. If the relationship between {\ccom} and {\cbri} still appears even after the environmental dependence is controlled, the properties of faint galaxies are probably affected by their adjacent bright galaxies in the {\vc}. For this, we define three distinct areas of the {\vc}, as shown in Figure~\ref{gp}, using the gravitational potential index: A1, A2, and A3. A1 is the center and A3 is the outermost part of the {\vc}. These areas were empirically selected so that the dependence of {\ccom} on {\gpi} is statistically negligible in each area. The selected criteria are {\gpi} = 0.47 (between A1 and A2) and {\gpi} = 0.25 (between A2 and A3). The boundary between A1 and A2 roughly corresponds to the virial radii of M87 and M49 as shown in Figure~\ref{gp}. In addition, the shape of the inner contour is similar to the shape of the outline of the central X-ray distribution \citep{boh94}. This indicates that the hot gas distribution follows the gravitational potential of the cluster. Since hot gas is known to be an important source of cluster environmental effects (i.e. ram pressure stripping), it is meaningful to use an environmental parameter which traces the distribution of hot gas. 

Figure~\ref{colcol_div} shows the plots of {\ccom} versus {\cbri} and 1/{\gpi} in the three areas. In all three areas, the red bright galaxies tend to have red faint companions like the result of the entire area, whereas {\ccom} do not significantly depend on the potential indices in any area. The overall trend in A1 is not very different from that over the entire area, but the linear fit slope is shallower and the correlation is less significant ($\emph{S}\sim1.75$ and $\emph{cc}\sim$0.27). On the other hand, although the number of points is much smaller than that in A1, the result for A2 shows a clearer color correlation between bright galaxies and their faint companions: $\emph{S}\sim2.73$ and $\emph{cc}\sim0.50$. The trend in A3 is also similar, but its $\emph{S}$ is approximately 1.59 and $\emph{cc}$ is approximately 0.44. The slope \emph{a}, bootstrap uncertainty, slope significance to the bootstrap uncertainty \emph{S} and \emph{cc} in each panel of Figures~\ref{colcol_full} and ~\ref{colcol_div} are displayed in Table~\ref{tres}. 

In every case, the slope significance in the {\ccom} versus 1/{\gpi} plot is smaller than 1.2$\,\sigma$; that is, the cluster-scale environmental dependence is almost negligible in all divided areas. On the other hand, the slope significance in the {\ccom} versus {\cbri} plot is larger than $1.5\,\sigma$. Particularly, the \emph{S} value in A2 is as large as $2.7\,\sigma$ ($\emph{cc}\sim0.50$), which indicates that {\ccom} obviously depends on {\cbri} in this area. That is, it appears that the {\ccom} versus {\cbri} relation slope is very small and marginal in the central area (A1), but it becomes larger and more significant in the outer area (A2). In A3, the slope and its significance seem to be smaller and weaker than those in A2, but there are too few points in A3. Statistically, the difference in the slope between A2 and A3 is not significant, although, neither is that between A1 and A3.   

\section{DISCUSSION}\label{dc}
The galactic conformity is very marginal in the central area of the {\vc} (A1), which seems to be similar to the result in {\mc} from \citet{lee14}. Since the analysis in \citet{lee14} was limited to the central area within 500 kpc radius, the two clusters (Virgo and {\mc}) commonly show only weak signals of small-scale conformity in their central areas ($\sim2\,\sigma$ significance). On the other hand, the signal of small-scale conformity in A2 appears to be more obvious, showing significance close to $3\,\sigma$. In the case of the outermost area A3, as mentioned in the previous section, the signal is very weak due to the insufficient size of the sample. It is noted that the correlation coefficient in A3 is as large as 0.44, which is similar to the value in A2, while that in A1 is only 0.27. This is not a decisive evidence of small-scale conformity in A3, but it shows a possibility that a clearer signal of conformity may be detected if a larger sample is available. In this paper, however, because the uncertainty for the result in A3 is too large, we will focus on the results in A1 and A2. These results show that small-scale conformity is weak or somewhat ambiguous in the inner region (roughly within the virial radius), while it seems to be more obvious in the outer region of the {\vc}.

Note that, in Section~\ref{defgal}, the average standard deviation of color indices in faint companions around a given bright galaxy is 0.116. This value is only approximately one fourth of the {\ccom} distribution over the entire sample, which ranges from 0.25 to 0.70. This small value of the average standard deviation of color indices implies that the faint companion galaxies in a possible group tend to have similar properties rather than have randomly mixed properties, which is consistent with the expected small-scale conformity.

\citet{lee14} suggested three scenarios regarding the origin of the small-scale galactic conformity in a cluster. The first is the ``tidal dwarfs'' scenario, in which the massive galaxies are partially torn out by tidal interactions with many other galaxies (harassment) or the cluster potential itself, forming many low-mass galaxies that may be the faint companions we observe. The properties of these low-mass galaxies must be similar to their host massive galaxy because they have been a single galaxy. If this scenario is correct, the conformity may be detected better in the inner region than in the outer region of a cluster, because the tidal force by a cluster is stronger in the inner region and more numerous massive galaxies exist in the inner region of a typical cluster, which may cause stronger harassment. However, this does not agree with our results that show an exactly opposite trend. Furthermore, the feasibility of the scenario itself needs to be confirmed, particularly regarding how efficiently such tidal dwarfs can be formed in a cluster.

The second is the ``dwarf capture'' scenario. Here, the massive galaxies are thought to capture low-mass galaxies near them and form a new small-scale galaxy system \citep{bas98,din00,ber03}. Within the new system, which may corresponds to our possible groups, the properties of the bright host galaxy and its new faint satellites may come to be similar because they share the same small-scale environment and may interact with each other. However, this may be less likely to happen as the cluster-centric distance decreases, because the relative velocities between galaxies become higher. On the other hand, the capturing events may occur more easily in the outer region, where the relative velocities are moderate, and thus this scenario seems to be consistent with our results to some extent. However, one critical problem is the timescale for the captured low-mass galaxies to sufficiently interact with the massive galaxy. To form the small-scale conformity in this way, the time scale of galaxy interaction should be much shorter than the timescale of galaxy infall. Another problem is that the galaxy number density decreases with increasing cluster-centric distance, which means that the probability of capture also decreases. For the second scenario to be shown to be plausible, the effects of these two aspects should be made clear.   

Last is the ``vestige of infallen groups'' scenario, in which the small-scale conformity in a cluster may be the vestige of infallen galaxy groups, but it disappears as the groups come closer to the center of the cluster. This scenario is based on the hierarchical large-scale structure formation, in which a galaxy cluster grows by merging smaller systems such as galaxy groups \citep{pre74,got75,whi78}. If a galaxy group falls into a galaxy cluster, the small-scale conformity may survive for some time after the infall event, but as time goes, the conformity will be continuously weakened by cluster environmental effects. If the galaxies closer to the cluster center are those that have fallen into the cluster longer ago, this scenario predicts stronger small-scale conformity in the outer region than near the center. Hence, this scenario is quite consistent with our results. If this scenario is correct, the small-scale conformity in a cluster may depend on the dynamical stage of the cluster. For instance, dynamically younger clusters may show more obvious small-scale conformity, because they have more newly infallen groups, while dynamically relaxed clusters may hardly show small-scale conformity. Since the Virgo cluster is known to be dynamically young, more studies of galaxy clusters in various dynamical stages will be very helpful to confirm the origin of the small-scale conformity in a cluster.

\begin{deluxetable}{cr @{.} lr @{.} lr @{.} lr @{.} l}
\tablenum{2} \tablecolumns{9} \tablecaption{Results using various magnitude cuts \label{magcut}} \tablewidth{0pt}
\tablehead{& \multicolumn{4}{c}{A1} & \multicolumn{4}{c}{A2} \\
\hline
Magnitude cut & \multicolumn{2}{c}{Significance} & \multicolumn{2}{c}{Correlation} & \multicolumn{2}{c}{Significance} & \multicolumn{2}{c}{Correlation} \\
& \multicolumn{2}{c}{(\emph{S})} & \multicolumn{2}{c}{Coefficient (cc)} & \multicolumn{2}{c}{(\emph{S})} & \multicolumn{2}{c}{Coefficient (cc)}}
\startdata
$-17.0$ & $0$ & 397 & $0$ & 082 & $2$ & 686 & $0$ & 482 \\
$-17.5$ & $0$ & 505 & $0$ & 140 & $2$ & 755 & $0$ & 442 \\
$-18.0$ $^{(i)}$ & $1$ & 746 & $0$ & 273 & $2$ & 733 & $0$ & 498 \\
$-18.5$ & $2$ & 140 & $0$ & 337 & $2$ & 607 & $0$ & 489 \\
$-19.0$ & $1$ & 954 & $0$ & 240 & $2$ & 701 & $0$ & 551 \\
\enddata
\tablecomments{$(i)$ Original magnitude cut.}
\end{deluxetable}

Finally, we discuss several caveats in this study. First, we applied a magnitude cut ($M_r \le -18$) to select possible group-central bright galaxies, but this simplistic criterion may not work perfectly. That is, neither all galaxies brighter than -18 magnitude in the $r$-band were central galaxies of groups, nor were all galaxies fainter than that satellite galaxies. Since it is technically very difficult to resolve this issue, we rely on a statistical approach for better selection. \citet{lan15} investigated the galaxy luminosity functions of groups and clusters at $z\sim$ 0 over a very wide range of luminosity and halo mass using the SDSS DR7 data. Their results show that the field luminosity function is dominated by the contributions of group-central galaxies at magnitudes brighter than approximately -18 mag in the $r$-band, while it is dominated by satellite galaxies at magnitudes fainter than that (Figure 7 in \citealt{lan15}). Such variation is reflected in the change of the luminosity function slope around $M_r \sim -18$: the luminosity functions of central and satellite galaxies have different shapes. Hence, our selection of bright galaxies is statistically appropriate, although it is not perfect. 

We speculate on the effect of falsely selected central galaxies and satellite galaxies. In the case of a falsely selected central galaxy, even though it is not a genuine central galaxy but a satellite in a larger group, the falsely selected group (the falsely selected central and its faint companions) may be a part of the larger group. Thus, if conformity exists in the larger group, it may also appear in the falsely selected group, which implies that the falsely selected central may not affect the result significantly. In the case of a falsely selected satellite galaxy, it may weaken the conformity because it is not a genuine member of the possible group we selected. However, such an effect may not be significant because we use the mean color of faint companions, a large fraction of which is expected to be genuine satellites. To examine our speculation, we tested various selections of magnitude cut in A2: from -17.0 mag (more falsely selected central galaxies) to -19.0 (more falsely selected satellites). The results are consistent with the original result as shown in Table~\ref{magcut}. That is, the falsely selected central galaxies and satellites seem not to affect the result significantly. 

Second, in our definition of possible groups in the Virgo cluster, some faint galaxies were regarded as the companions of multiple bright galaxies at the same time, which may affect the signal of the small-scale conformity. We speculate that such an effect tends to weaken the apparent small-scale conformity by confusing the small-scale membership, rather than to strengthen it by mere coincidence. Such an effect may be strongest in A1, because the galaxy number density is highest in the cluster center. We also examined various selections of magnitude cut (from -17.0 to -19.0 mag) in A1, to examine the effects of satellite duplication. Although the change of magnitude cut does not significantly affect the result as discussed in the previous paragraph, the smaller (brighter) magnitude cut reduces the number of possible groups and thus reduces the number of duplicated satellites. The results show measurable variations: no dependence for -17.0 and -17.5 mag cuts (more duplication), while more obvious trends for -18.5 and -19.0 mag cuts (less duplication) as listed in Table~\ref{magcut}. This indicates that satellite duplication may have weakened the intrinsic conformity in A1, by increasing the number of false satellites in each group. Thus, the conformity in the cluster center may be stronger than our measurement. However, it appears to be still weaker than the conformity in the outer region, which implies that our conclusion may not change.

Third, the {\gpi} parameter considers only the gravitation of M87 and M49, despite the existence of more relatively massive galaxies in the {\vc}, such as M60 or M86. This is because we focused on the most massive galaxies that significantly affect the distribution of the intracluster medium in the {\vc}. As shown in Figure~\ref{gp}, the hot gas distribution (traced by the X-ray map) is determined mainly by M87 and M49, whereas the importance of the other galaxies seems to be relatively weak. This indicates that the {\gpi} parameter we defined may reflect well the effect of the cluster hot gas (i.e., ram pressure stripping). 

Fourth, although the statistical uncertainty is very large, the slope of the {\ccom} versus {\cbri} relation in A3 appears to be smaller than that in A2. If the small-scale conformity in the outermost area is actually weaker than that in the intermediate area, it would indicate that the conformity has formed during the infall process, not in-situ; that is, it would support the ``dwarf capture" scenario rather than the ``vestige of infallen groups" scenario. This is not confirmed in this paper, and moreover it may not be easy to check this in other clusters, because the galaxy number density in such an outermost area of a typical cluster is low. A stacking analysis for multiple clusters could be tried, but it too would be challenging.

\section{CONCLUSION}\label{cc}

We carried out a study on the {\vc} to understand the small-scale conformity between bright galaxies ($M_r \le -18$) and their faint companions ($M_r > -18$) in a cluster using the EVCC and the SDSS DR12 data. In the entire area, the luminosity-weighted mean color of faint companion galaxies {\ccom} depends not only on the color of their bright central galaxies {\cbri}, but also on the cluster gravitational potential index {\gpi}. To examine the genuine {\ccom} versus {\cbri} relations when the cluster-scale environmental effect is controlled, we defined three sub-areas (A1, A2, and A3), in which the dependence of {\ccom} on {\gpi} is almost negligible. The small-scale conformity appears to be obvious in A2, whereas it does not exist or is very marginal in A1. In A3, the result is unclear due to the insufficient sample size. Therefore, we conclude that the small-scale conformity of the {\vc} galaxies is very weak in the inner region but it is more obvious in the outer region. This result is consistent with the ``vestige of infallen groups'' scenario, although the ``dwarf capture" scenario cannot be rejected explicitly. To confirm the origin of the small-scale conformity in a cluster, more studies of galaxy clusters in various evolutionary stages are required. Owing to the fact that the Virgo cluster and the previously studied {\mc} are both thought to be dynamically young, case studies for dynamically relaxed clusters with reliable spectroscopic membership information will be particularly valuable.  

\acknowledgements

H.J. acknowledges support from the Basic Science Research Program through the National Research Foundation of Korea (NRF), funded by the Ministry of Education (NRF-2013R1A6A3A04064993).

\end{document}